\documentclass[sn-mathphys]{sn-jnl}

\jyear{2021}%

\usepackage{subfigure}
\graphicspath{{figures/}}
\usepackage{siunitx}
\usepackage{makecell}
\usepackage{lineno}
\usepackage[export]{adjustbox}
\usepackage{upgreek}
\usepackage{color}

\theoremstyle{thmstyleone}%
\theoremstyle{thmstyletwo}%

\theoremstyle{thmstylethree}%

\begin{document}

\title[In-orbit background simulation of a type-B CATCH satellite]{In-orbit background simulation of a type-B CATCH satellite}


\author[1,2]{\fnm{Jingyu} \sur{Xiao}}

\author[1]{\fnm{Liqiang} \sur{Qi}}\email{qilq@ihep.ac.cn}

\author[1,2]{\fnm{Shuang-Nan} \sur{Zhang}}\email{zhangsn@ihep.ac.cn}

\author[1]{\fnm{Lian} \sur{Tao}}

\author[1]{\fnm{Zhengwei} \sur{Li}}

\author[1]{\fnm{Juan} \sur{Zhang}}

\author[1]{\fnm{Xiangyang} \sur{Wen}}

\author[1]{\fnm{Qian-Qing} \sur{Yin}}

\author[1]{\fnm{Yanji} \sur{Yang}}

\author[3]{\fnm{Qingcui} \sur{Bu}}

\author[1]{\fnm{Sheng} \sur{Yang}}

\author[1]{\fnm{Xiaojing} \sur{Liu}}

\author[1,2]{\fnm{Yiming} \sur{Huang}}

\author[4]{\fnm{Wen} \sur{Chen}}

\author[4]{\fnm{Yong} \sur{Yang}}

\author[4]{\fnm{Huaqiu} \sur{Liu}}

\author[4]{\fnm{Yibo} \sur{Xu}}

\author[1,2]{\fnm{Shujie} \sur{Zhao}}

\author[1,5]{\fnm{Xuan} \sur{Zhang}}

\author[1,2]{\fnm{Panping} \sur{Li}}

\author[1,2]{\fnm{Kang} \sur{Zhao}}

\author[1,2]{\fnm{Ruican} \sur{Ma}}

\author[1,2]{\fnm{Qingchang} \sur{Zhao}}

\author[6]{\fnm{Ruijing} \sur{Tang}}

\author[1,5]{\fnm{Jinhui} \sur{Rao}}

\author[1,7]{\fnm{Yajun} \sur{Li}}

\affil[1]{\orgdiv{Key Laboratory of Particle Astrophysics, Institute of High Energy Physics}, \orgname{Chinese Academy of Sciences}, \orgaddress{\city{Beijing}, \postcode{100049}, \country{China}}}

\affil[2]{\orgdiv{University of Chinese Academy of Sciences}, \orgname{Chinese Academy of Sciences}, \orgaddress{\city{Beijing}, \postcode{100049}, \country{China}}}

\affil[3]{\orgdiv{Institut für Astronomie und Astrophysik, Kepler Center for Astro and Particle Physics}, \orgname{Eberhard Karls Universität}, \orgaddress{\street{Sand 1}, \city{Tübingen}, \postcode{72076}, \country{Germany}}}

\affil[4]{\orgdiv{Innovation Academy for Microsatellites of Chinese Academy of Sciences}, \orgaddress{\city{Shanghai}, \postcode{200135}, \country{China}}}

\affil[5]{\orgdiv{Nanchang University}, \orgaddress{\city{Nanchang}, \postcode{330031}, \country{China}}}

\affil[6]{\orgdiv{Beijing Jiaotong University}, \orgaddress{\city{Beijing}, \postcode{100044}, \country{China}}}

\affil[7]{\orgdiv{Zhengzhou University}, \orgaddress{\city{Zhengzhou}, \postcode{450001}, \country{China}}}

\abstract{The Chasing All Transients Constellation Hunters (CATCH) space mission plans to launch three types of micro-satellites (A, B, and C). The type-B CATCH satellites are dedicated to locating transients and detecting their time-dependent energy spectra. A type-B satellite is equipped with lightweight Wolter-I X-ray optics and an array of position-sensitive multi-pixel Silicon Drift Detectors. To optimize the scientific payloads for operating properly in orbit and performing the observations with high sensitivities, this work performs an in-orbit background simulation of a type-B CATCH satellite using the Geant4 toolkit. It shows that the persistent background is dominated by the cosmic X-ray diffuse background and the cosmic-ray protons. The dynamic background is also estimated considering trapped charged particles in the radiation belts and low-energy charged particles near the geomagnetic equator, which is dominated by the incident electrons outside the aperture. The simulated persistent background within the focal spot is used to estimate the observation sensitivity, i.e.\ 4.22~$\times$~10$^{-13}$~erg~cm$^{-2}$~s$^{-1}$ with an exposure of 10$^{4}$~s and a Crab-like source spectrum, which can be utilized further to optimize the shielding design. The simulated in-orbit background also suggests that the magnetic diverter just underneath the optics may be unnecessary in this kind of micro-satellites, because the dynamic background induced by charged particles outside the aperture is around 3 orders of magnitude larger than that inside the aperture.}  


\maketitle

\section{Introduction}\label{sec1}

The Chasing All Transients Constellation Hunters (CATCH) space mission was proposed in 2019. It aims to simultaneously monitor abundant transients in the universe and depict a dynamic view of the universe with the multi-wavelength and multi-observable measurement of X-rays~\cite{li2023catch}. The mission plans to form an intelligent X-ray constellation with 126 satellites around 2030. It enables simultaneously the all-sky monitoring, follow-up observations, and multi-observable measurement of X-rays with a flexible Field-of-View (FoV). The constellation includes three types of micro-satellites with each placing its own particular emphasis on the observation capability. The type-B CATCH satellites are mainly used for locating the transients, and performing the spectral and timing measurements in the energy band between 0.3~keV and 10~keV with high sensitivities. A type-B satellite is equipped with a lightweight Wolter-I focusing mirror and an array of position-sensitive multi-pixel Silicon Drift Detectors (SDDs) to measure the incident positions, energy, and arrival time of X-rays.

The in-orbit background simulation is essential in the development of X-ray telescopes before launch. It can be used to estimate the observation sensitivity of the instrument, optimize the instrument design, evaluate the onboard storage requirements, and infer the degradation of detector performances due to the radiation damage. Such studies have been performed in the space missions of Suzaku~\cite{yamaguchi2006background}, Swift~\cite{pagani2007characterization}, eROSITA~\cite{perinati2012radiation,tenzer2010geant4,weidenspointner2008application}, ATHENA~\cite{fioretti2016monte}, LOFT~\cite{campana2013background}, Insight-HXMT~\cite{xie2015simulation,xie2018study,zhang2020comparison}, and EP~\cite{zhang2022estimate}. The Geant4 toolkit, developed by CERN, is now widely used in space science for the in-orbit background simulation~\cite{agostinelli2003geant4,allison2006geant4,allison2016recent}. It is based on the Monte Carol method and has multiple advantages against a deterministic algorithm. It can simulate the passage of particles through complex geometries and materials with the built-in physics library of the interaction between particles and matters. Thus, the Geant4 toolkit is chosen in this work to simulate the in-orbit background of a type-B CATCH satellite. 

In this manuscript, Section 2 introduces the space radiation environment for the in-orbit background simulation of a type-B CATCH satellite. Section 3 presents the mass modeling of the satellite, including the geometry, physics processes, and primaries generation. Section 4 shows the simulated results including the persistent and dynamic in-orbit background, and the sensitivity based on them. Finally, the summaries and conclusions are given in Section 5.
 
\section{Radiation environment}\label{sec2}

\begin{figure}[ht]%
\centering
\includegraphics[width=0.92\textwidth]{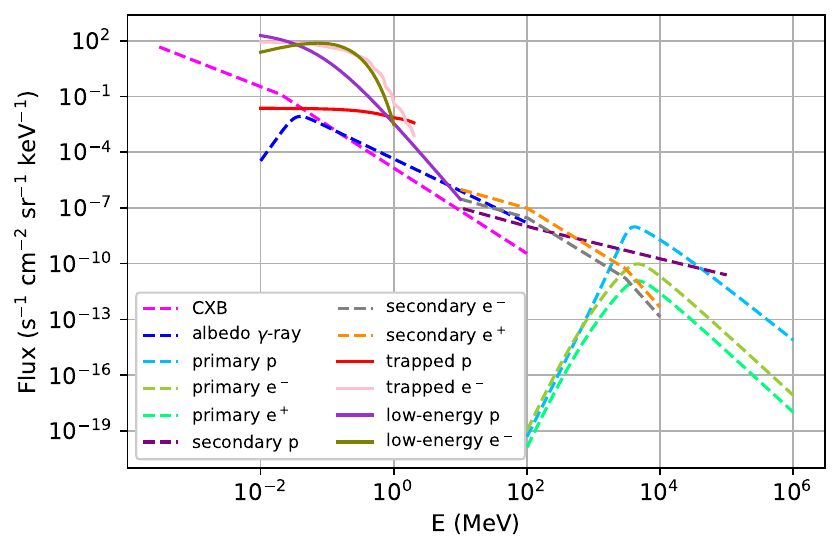}
\vspace{0.4cm}
\caption{Energy spectra of all components in the space radiation environment, which serve as the input to the background simulation of a type-B CATCH satellite in this work}\label{spectrumall}
\end{figure}

The Type-B CATCH satellite is in the low-Earth orbit (LEO). Its orbital altitude and inclination are 550~km and $\ang{29;;}$~\cite{li2023catch}, respectively. The radiation environment of the satellite mainly includes the photon and cosmic-ray background. The energy spectra of all components in the radiation environment are calculated with the analytical formulas often used by other LEO satellites and are plotted in Figure~\ref{spectrumall}. They are used as the input to the background simulation and are detailed in this section.

\subsection{Photon background}

The photon background includes the cosmic X-ray diffuse background (CXB) and the atmospheric albedo $\gamma$-ray. In the soft X-ray band, the main contribution to the CXB comes from the thermal emission from the hot gas in the Galaxy. In the hard X-ray band, the CXB originates from many unresolved X-ray sources outside the Galaxy~\cite{gilli2007synthesis,lahav1993significant,hickox2019resolving,gruber1999spectrum}. The atmospheric albedo $\gamma$-ray is the reflection of $\gamma$-ray by the atmosphere to the satellite~\cite{2007hsaa.book.....Z}, and the $\gamma$-ray is the product of the interaction between primary cosmic-ray particles and the Earth's atmosphere. 

\subsubsection{CXB}

A broken power-law spectrum is used to represent the CXB in the energy range between 0.3~keV and 10$^{5}$~keV as follows~\cite{gehrels1992instrumental},
\begin{equation}
    F\left(E\right)=\left\{\begin{array}{ll}
    0.54\times 10^{-3}\times \left(\frac{E}{1000}\right)^{-1.4}, & E<20\:\mathrm{keV} \\
    0.0117\times 10^{-3}\times \left(\frac{E}{1000}\right)^{-2.38}, & 20\:\mathrm{keV}\leq E<100\:\mathrm{keV} \\
    0.014\times 10^{-3}\times \left(\frac{E}{1000}\right)^{-2.3}, & E\geq 100\:\mathrm{keV}
    \end{array}\right.,\label{equ1}
\end{equation}
where $F$ is in units of $\rm counts\:cm^{-2}\:s^{-1}\:sr^{-1}\:keV^{-1}$ and $E$ is in units of keV.

\subsubsection{Albedo \texorpdfstring{$\gamma$}{Lg}-ray}

The albedo $\gamma$-ray spectrum adopted in this work is as follows~\cite{campana2013background},
\begin{equation}
    F\left(E\right)=\frac{0.0148}{\left(\frac{E}{33.7\:\mathrm{keV}}\right)^{-5}+\left(\frac{E}{33.7\:\mathrm{keV}}\right)^{1.72}},\label{equ2}
\end{equation}
where $F$ is in units of $\rm counts\:cm^{-2}\:s^{-1}\:sr^{-1}\:keV^{-1}$ and the considered energy range is between 10~keV and 10$^{5}$~keV. The albedo $\gamma$-ray flux depends on the relative position between the satellite and the Earth. Assuming a type-B CATCH satellite points to the zenith, it is only affected by the albedo $\gamma$-ray within a certain opening angle. The opening angle $\theta$ is determined by $\theta=2\arcsin{(R_\textup{Earth}/(R_\textup{Earth}+h))}$, where $R_\textup{Earth}$ is the Earth's radius and $h$ the satellite's orbital height.

\subsection{Cosmic-ray background}

The universe is full of various cosmic rays. Primary cosmic rays are high-energy particles from the extraterrestrial. Secondary cosmic rays are produced when primary cosmic rays collide with the atomic nuclei in the Earth's atmosphere~\cite{2007hsaa.book.....Z}. Satellites are also affected by energetic charged particles trapped in the Van Allen radiation belts~\cite{ginet2014ae9}. The flux enhancements of low-energy protons and low-energy electrons near the geomagnetic equator were observed~\cite{petrov2008energy,grigoryan2008spectral}.

\subsubsection{Primary cosmic rays}

The energy spectra of primary cosmic rays can be described by a unified formula for protons, electrons, and positrons as follows~\cite{mizuno2004cosmic},
\begin{align}
F\left(E\right)&=F_\textup{U}\left(E+Ze\phi\right)\times F_\textup{M}\left(E,M,Z,\phi\right)\times C\left(R,h,\theta_\textup{M}\right)\nonumber \\
&=A\left[\frac{R\left(E+Ze\phi\right)}{\mathrm{GV}}\right]^{-a}\times \frac{\left(E+Mc^{2}\right)^{2}-\left(Mc^{2}\right)^{2}}{\left(E+Ze\phi+Mc^{2}\right)^{2}-\left(Mc^{2}\right)^{2}}\times \frac{1}{1+\left(R/R_\textup{cut}\right)^{-r}},\label{equ:FE}
\end{align}
where $R$ is the magnetic rigidity, $R_\textup{cut}$ the cutoff rigidity, and $\phi$ the solar modulation potential~\cite{gleeson1968solar}. $F$ in Equation~\ref{equ:FE} is in units of $\rm counts\:cm^{-2}\:s^{-1}\:sr^{-1}\:keV^{-1}$ and $E$ is in units of keV. The parameters for different particles are summarized in Table~\ref{paraprimary}. The considered energy range is between 10$^{5}$~keV and 10$^{9}$~keV.
\begin{table}[ht]
\begin{center}
\begin{minipage}{\textwidth}
\caption{Summary of the parameters in the energy spectra of primary cosmic rays }\label{paraprimary}%
\centering
\resizebox{\textwidth}{!}{
\begin{tabular}{@{}lcccccccc@{}}
\toprule
Particle & \makecell[c]{$A$\\($\rm counts\:cm^{-2}\:s^{-1}\:sr^{-1}\:keV^{-1}$)}  & \makecell[c]{$Mc^{2}$\\(MeV)} & \makecell[c]{$Ze$\\(e)} & $a$ & $r$ & \makecell[c]{$\phi$\\(MV)} & $\theta_\textup{M}$ & $R_\textup{cut}$\\
\midrule
p   &  $23.9\times 10^{-7}$  &  923  &  1   & 2.83 &  12  &  900  &  0.7 & 4.32\\ 
$e$ &  $0.65\times 10^{-7}$  & 0.511 &  1   & 3.30 &  6   &  900  &  0.7 & 4.32\\ 
$e^{+}$ &  $0.08\times 10^{-7}$  & 0.511 &  1   & 3.30 &  6   &  900  &  0.7 & 4.32 \\ 
\botrule
\end{tabular}}
\end{minipage}
\end{center}
\end{table}

\subsubsection{Secondary cosmic rays}

The energy spectra of secondary cosmic rays can be represented by a broken power-law as follows~\cite{mizuno2004cosmic},
\begin{equation}
    F\left(E\right)=\left\{\begin{array}{ll}
    A\left(\frac{E}{100\:\mathrm{MeV}}\right)^{-1}, & 10^{4}\:\mathrm{keV}\leq E< 10^{5}\:\mathrm{keV} \\
    A\left(\frac{E}{100\:\mathrm{MeV}}\right)^{-a}, & 10^{5}\:\mathrm{keV}\leq E< E_\textup{bk} \\
    A\left(\frac{E_\textup{bk}}{100\:\mathrm{MeV}}\right)^{-a}\left(\frac{E}{E_\textup{bk}}\right)^{-b}, & E\geq E_\textup{bk} 
    \end{array}\right.,\label{equ:F}
\end{equation}
where $F$ is in unit of $\rm counts\:cm^{-2}\:s^{-1}\:sr^{-1}\:keV^{-1}$ and the parameters are related to the geomagnetic latitude $\theta_\textup{M}$ (see Table~\ref{parasecondary}). The considered energy range for protons is between 10$^{4}$~keV and 10$^{8}$~keV. The considered energy range for electrons and positrons is between 10$^{4}$~keV and 10$^{7}$~keV. 

\begin{table}[ht]
\begin{center}
\begin{minipage}{\textwidth}
\caption{Summary of the parameters in the energy spectra of secondary cosmic rays}\label{parasecondary}%
\centering
\resizebox{\textwidth}{!}{
\begin{tabular}{@{}lccccccc@{}}
\toprule
Particle & \makecell[c]{$A$\\($\rm counts\:cm^{-2}\:s^{-1}\:sr^{-1}\:keV^{-1}$)}  & \makecell[c]{$E_\textup{bk}$\\(GeV)} & $a$ & $b$ & $\theta_\textup{M}$\\
\midrule
p   &  $0.1\times 10^{-7}$  & 600 & 0.87 & 2.53 & 0.2-0.3 \\ 
$e$ &  $0.3\times 10^{-7}$  & 3   & 2.2  &  4.0 &  0-0.3 \\ 
$e^{+}$ &  $0.99\times 10^{-7}$  & 3   & 2.2  &  4.0 &  0-0.3 \\
\botrule
\end{tabular}}
\end{minipage}
\end{center}
\end{table}

\subsubsection{Trapped protons and electrons}

The AP9/AE9 models (V1.5) describe the protons and electrons trapped by the Earth's radiation belts~\cite{ginet2014ae9}. According to the orbit of a type-B CATCH satellite, the average flux of trapped protons and electrons is obtained from SPENVIS~\footnote{https://www.spenvis.oma.be/}. Linear interpolation and extrapolation are used to obtain the spectra in the energy range between 10~keV and 2000~keV.

\subsubsection{Low-energy charged particles near the geomagnetic equator}

The spectrum of low-energy protons remains almost unchanged near the geomagnetic equator within the altitude between 500~km and 1000~km~\cite{petrov2008energy}. It can be described by a kappa function as follows~\cite{petrov2008energy}, 
\begin{equation}
    F\left(E\right)=A\left(1+\frac{E}{kE_\textup{0}}\right)^{-k-1},
\end{equation}
where $F$ is in units of $\rm counts\:cm^{-2}\:s^{-1}\:sr^{-1}\:keV^{-1}$, $A$~=~328, $k$~=~3.2, and $E_\textup{0}$~=~22~keV. The considered energy range is between 10~keV and $10^{4}$~keV.

The energy spectrum of low-energy electrons from 10~keV to 10$^{3}$~keV is a Maxwell function ($L<1.2$) as follows~\cite{grigoryan2008spectral},
\begin{equation}
    F\left(E\right)= A\frac{E}{E_\textup{0}}\exp\left(-\frac{E}{E_\textup{0}}\right),
\end{equation}
where $F$ is in units of $\rm counts\:cm^{-2}\:s^{-1}\:sr^{-1}\:keV^{-1}$, $A$~=~200, and $E_\textup{0}$~=~73~keV.

\section{Geant4 mass modeling}

In this section, the geometry of a type-B CATCH satellite, the corresponding physics processes of various background components, and primaries generation are described.

\subsection{Geometry}\label{sec3.1}

\begin{figure}[ht]%
\centering
\includegraphics[width=0.72\textwidth]{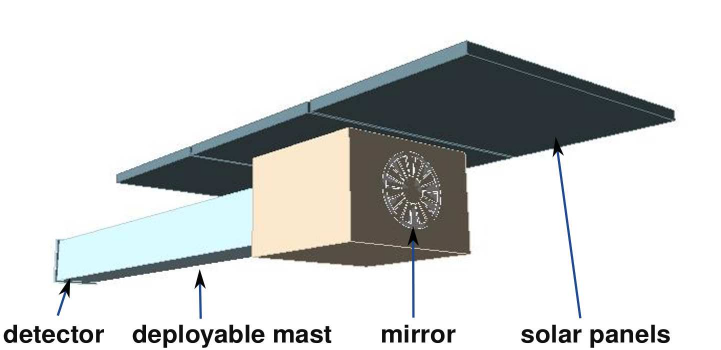}
\vspace{0.6cm}
\caption{Geometry model of a type-B CATCH satellite built in Geant4. The optics system is placed in the platform cabin, and the SDD array is at the back end of the satellite and connected to the platform through a deployable mast}\label{typeB}
\end{figure}

A type-B CATCH satellite as a whole looks like a rectangular box and the focal plane detector can be deployed 2 meters away from the platform cabin in orbit. The structure of it mainly includes a platform cabin, an optics system (30 layers of the Wolter-I type focusing mirror, filter films and spider wheel), solar panels, a deployable mast, and a detector system (a 256-pixel SDD array). To simplify the model and speed up the simulation, the platform cabin is replaced by Al with an equivalent density of 0.058~$\rm g\:cm^{-3}$. The solar panels are fixed on the platform cabin and unfolded, their material is replaced by Al. The deployable mast is also replaced by Al with a thickness of 0.01~mm. At the bottom, an Al plate is used to secure the mast. Figure~\ref{typeB} shows the geometry of a type-B CATCH satellite built in Geant4. The corresponding dimensions and parameters are summarized in Table~\ref{para0}.

\begin{table}[ht]
\begin{center}
\begin{minipage}{\textwidth}
\caption{Summary of the dimensions and parameters of the structures in a type-B CATCH satellite}\label{para0}%
\centering
\begin{tabular}{@{}lcc@{}}
\toprule
Item & Dimension & Material\\
\midrule
Platform Cabin  & 460\:mm$\times$320\:mm$\times$440\:mm & Al ($\rho=0.058\:\rm g\:cm^{-3}$)   \\
Solar Panel  & 800\:mm$\times$500\:mm$\times$30\:mm & Al ($\rho=2.699\:\rm g\:cm^{-3}$)   \\
Deployable Mast  & \makecell[c]{outline 200\:mm$\times$200\:mm$\times$1830\:mm\\effective thickness 0.01 mm} & Al ($\rho=2.699\:\rm g\:cm^{-3}$)  \\
Plate at the bottom of mast  & 220\:mm$\times$220\:mm$\times$15.38\:mm & Al ($\rho=2.699\:\rm g\:cm^{-3}$)  \\
\botrule
\end{tabular}
\end{minipage}
\end{center}
\end{table}

The focusing mirror is composed of the coaxial and confocal paraboloid and hyperboloid. Since the current version of Geant4 does not provide the class of two-sheet hyperboloids, this work uses the external class of G4Hyperboloid developed by Qi et al.~\cite{qi2020geant4} and the built-in G4Paraboloid class to describe the exact parabolic-hyperbolic geometry of the Wolter-I mirror shells. The filter at the entrance of the aperture comprises polyimide and Al. The spider wheel is used to fix the mirror shells with 18 spokes. The configurations of the optics are summarized in Table~\ref{para1}.

\begin{table}[ht]
\begin{center}
\begin{minipage}{\textwidth}
\caption{Summary of the configurations of Wolter-I optics in a type-B CATCH satellite}\label{para1}%
\centering
\begin{tabular}{@{}lc@{}}
\toprule
Item & Parameter\\
\midrule
Focal Length  & 2000\:mm  \\
Shell Number  & 30  \\
Shell-1 Intersection Radius $r_{0}$  & 97.50\:mm  \\
Shell Thickness  & $r_{0}\times$0.00177  \\
Layer Gap  & 0.2\:mm  \\
Paraboloid Length  & 200\:mm  \\
Hyperboloid Length  & 200\:mm  \\
Substrate Material  & Ni  \\
Coating Material  & Au  \\
Filter  & 400\:nm polyimide and 80\:nm Al  \\
Spider Wheel  & 40\:mm Inconel600 (18 spokes)  \\
\botrule
\end{tabular}
\end{minipage}
\end{center}
\end{table}

\begin{figure}[ht]%
\centering
\includegraphics[width=0.85\textwidth]{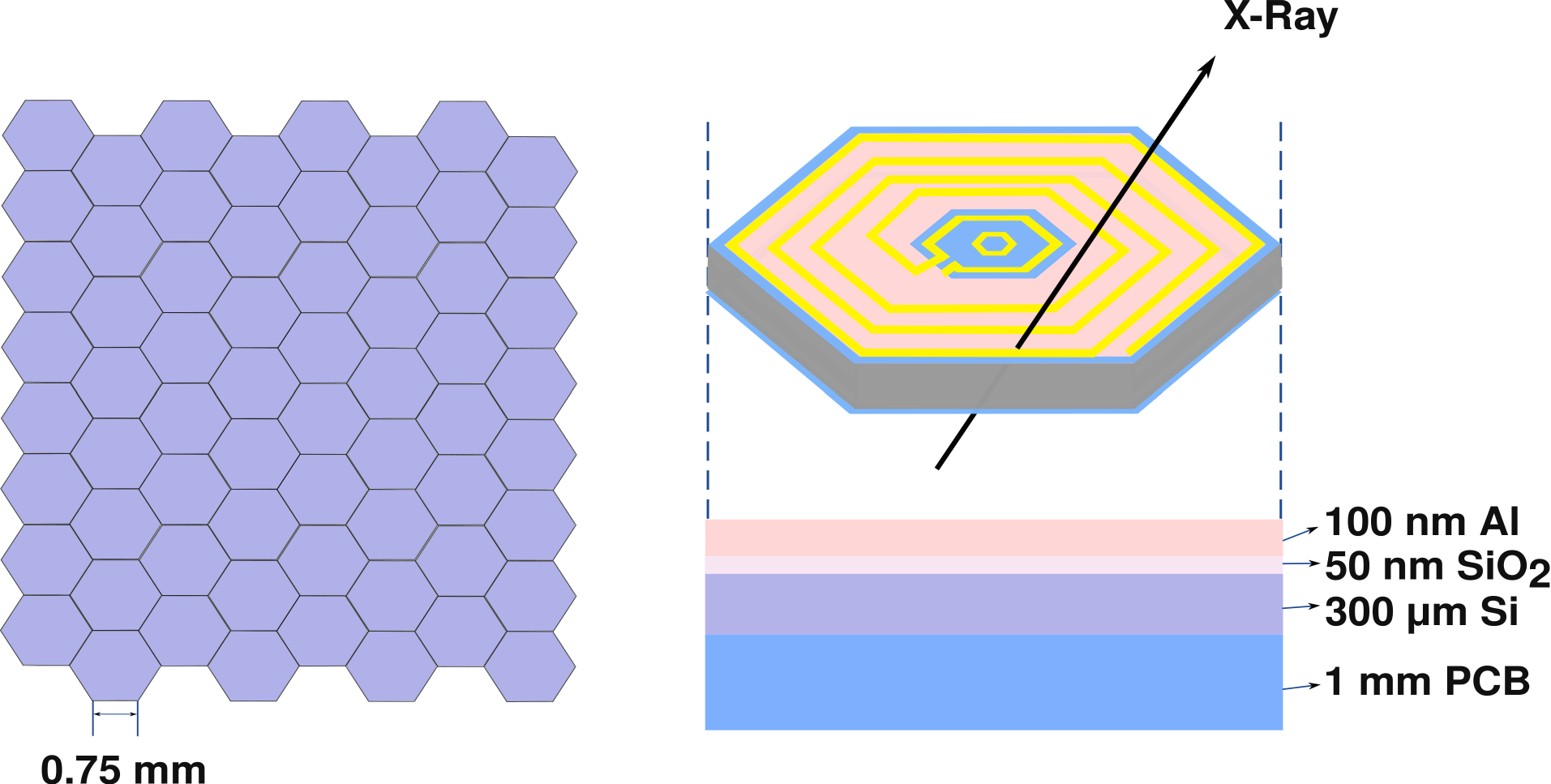}
\vspace{0.70cm}
\caption{Schematic view of one module of the SDD array in a type-B CATCH satellite (left panel). One pixel of the SDDs (right panel)}\label{SDD}
\end{figure}

\begin{figure}[ht]%
\centering
\includegraphics[width=0.55\textwidth]{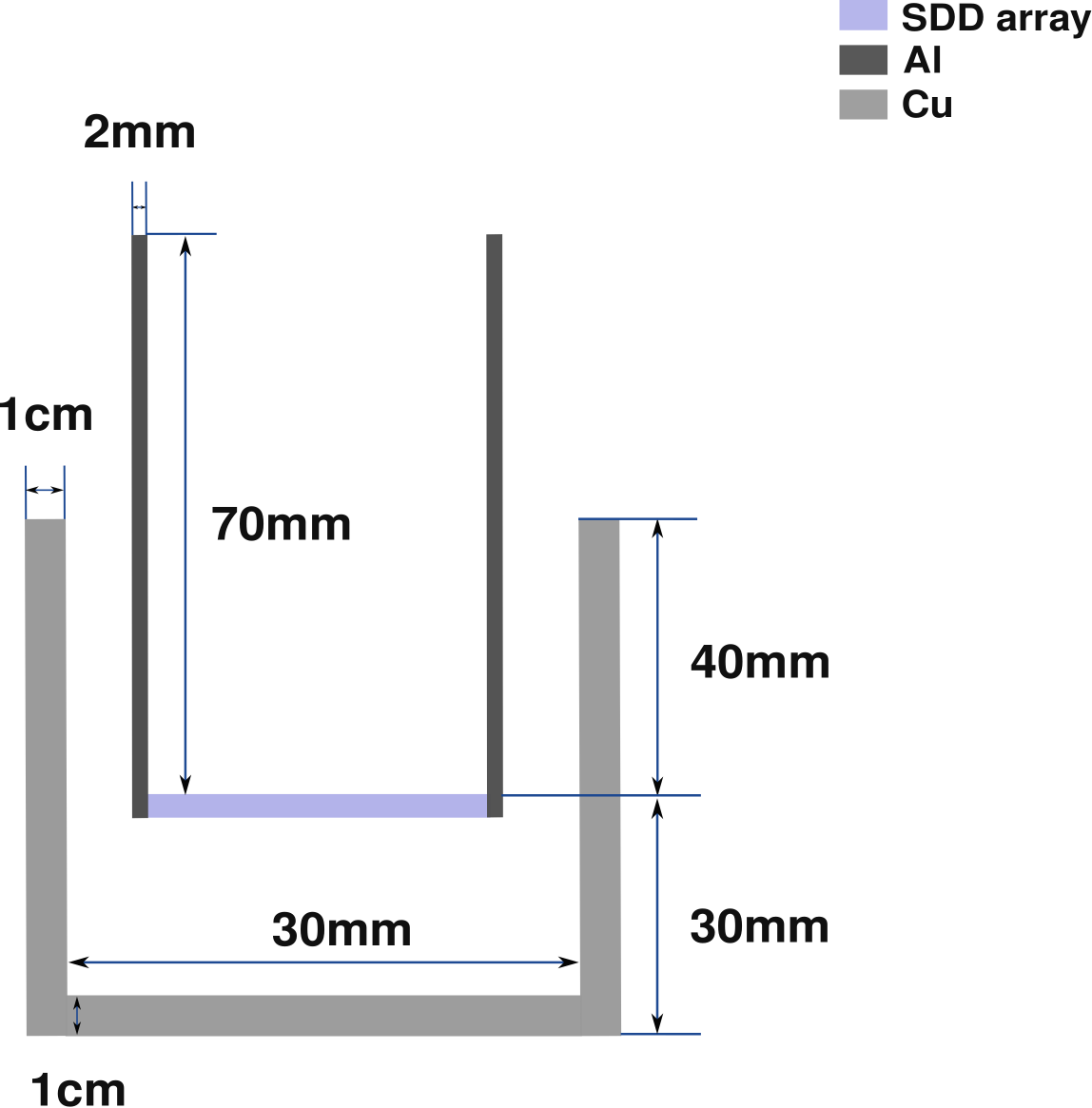}
\vspace{0.70cm}
\caption{Preliminary shielding design of the focal plane detector of a type-B CATCH satellite}\label{SDDshield}
\end{figure}

A type-B CATCH satellite uses an array of SDDs pixels at the focal plane (see Figure~\ref{SDD}). It contains 4 modules and each module has 64 pixels. Each pixel is hexagonal with a side length of 0.75 mm. The thickness of the depletion layer is 300~$\rm \mu m$ and the dead layer is 50~nm. To shield the optical light, an Al film with a thickness of 100~nm is coated on the surface of the detector. At the bottom, a ceramic printed circuit board is added to the model with an equivalent thickness of 1~mm. A preliminary shielding design of the detector includes 2~mm thick Al and 1~cm thick Cu, as shown in Figure~\ref{SDDshield}.

\subsection{Physics processes}\label{3.2}

The model provided by Geant4 can not describe many low-energy physical phenomena. Users have to develop the physics processes within the Geant4 framework to adapt to the application of X-ray telescopes, i.e.\ the total internal reflection of X-rays on the mirror surface and the funneling effect of low-energy charged particles through the Wolter-I type optics. Additionally, the interaction between particles and matters in the focal plane detector and surrounding materials uses the built-in physics processes of Geant4, i.e.\ mainly the standard and low-energy electromagnetic interactions, and hadron elastic and inelastic scattering.

\subsubsection{Mirror}

\begin{figure}[ht]%
    \begin{minipage}[b]{0.48\linewidth} 
        \centering 
        \includegraphics[width=1.12\textwidth,left]{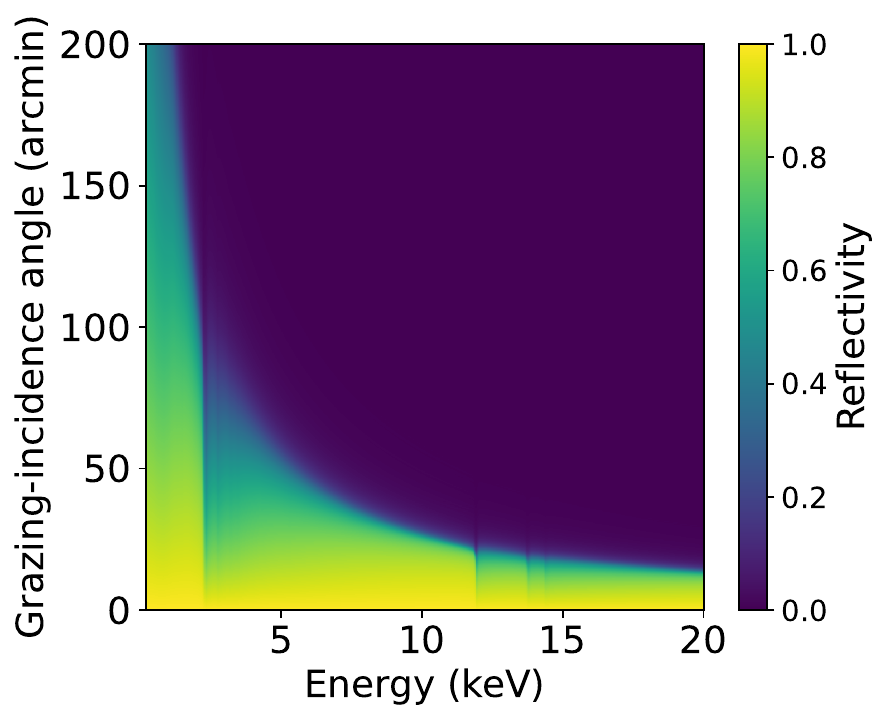} 
    \end{minipage} 
    \hspace{0.5cm}
    \begin{minipage}[b]{0.48\linewidth} 
        \centering 
        \includegraphics[width=0.95\textwidth]{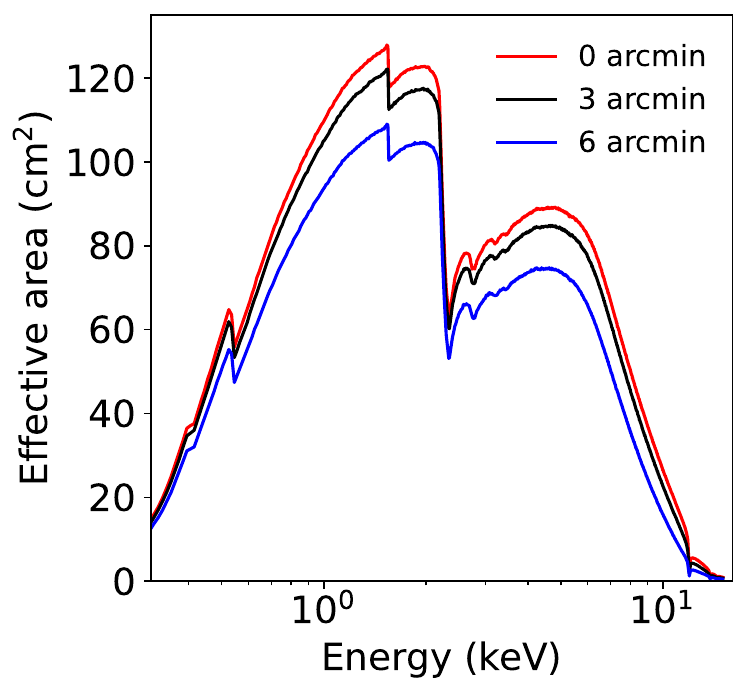} 
    \end{minipage}
    \vspace{0.1cm}
    \caption{Reflectivity as a function of the X-ray energy and grazing-incidence angle (left panel). The simulated effective area of the Wolter-I optics of a type-B CATCH satellite. Different curves represent incident X-rays at different off-axis angles (right panel)}
    \label{RandEA}
\end{figure}

\begin{figure}[ht]%
    \begin{minipage}[b]{0.48\linewidth} 
        \centering 
        \includegraphics[width=1.10\textwidth,left]{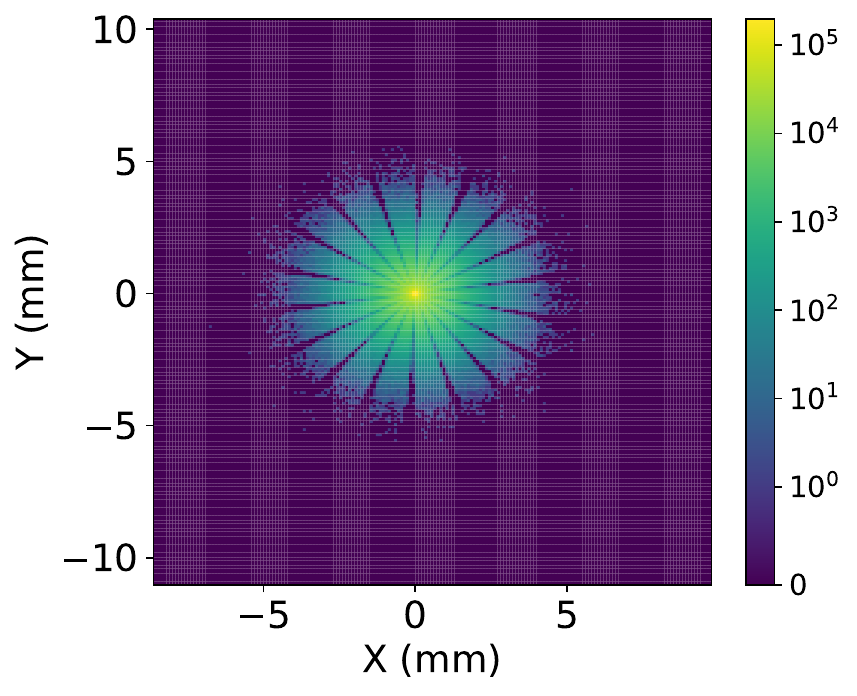} 
    \end{minipage} 
    \hspace{0.7cm}
    \begin{minipage}[b]{0.48\linewidth} 
        \centering 
        \includegraphics[width=0.98\textwidth]{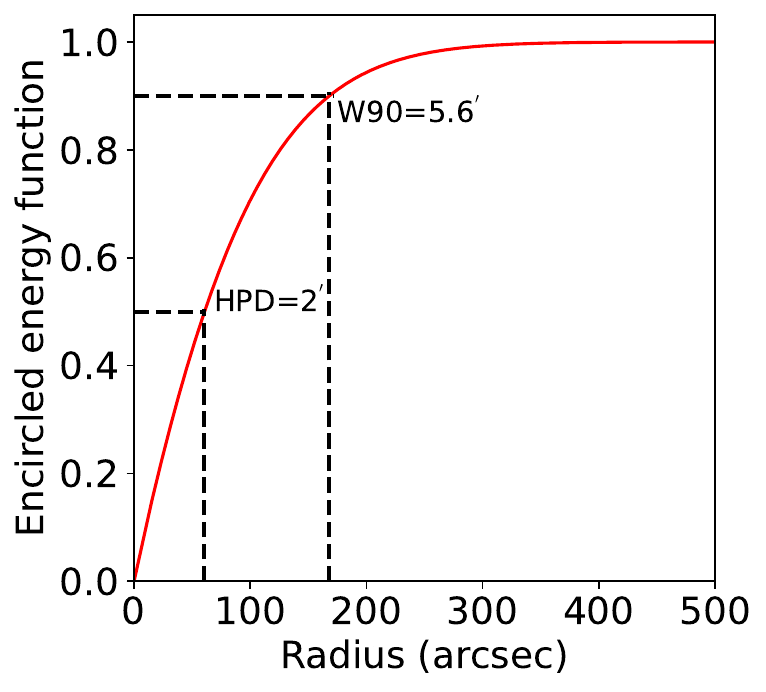} 
    \end{minipage}
    \vspace{0.1cm}
    \caption{Simulated on-axis PSF of the Wolter-I optics of a type-B CATCH satellite (left panel) and the corresponding on-axis EEF (right panel). The HPD and W90 are $\ang{;2.0;}$ and  $\ang{;5.6;}$, respectively}
    \label{PSFandEEF}
\end{figure}

An external physics process, G4XrayGrazingAngleScattering~\cite{qi2020geant4}, is implemented within the Geant4 simulation framework to represent the total internal reflection of X-rays on the mirror surface. It is invoked when an X-ray is incident on the surface of the pre-defined Wolter-I mirror shells. The reflectivity determines whether an X-ray is reflected or absorbed (the data are taken from~\footnote{https://www.cxro.lbl.gov/}). It is interpolated according to the energy and grazing-incidence angle of X-rays (see Figure~\ref{RandEA}). Since the actual mirror shape is imperfect, an X-ray photon does not follow specular reflection. In this work, the random surface theory is used to reproduce the imperfection of the mirror shape~\cite{spiga2013x}, i.e.\ a Gaussian function to generate a random deviation to the normal direction at the incident position and thus the direction of the reflected X-rays. The standard deviation of the Gaussian function can be adjusted to obtain the required imaging capabilities as the scientific requirement. Figure~\ref{PSFandEEF} shows the on-axis point spread function (PSF) and corresponding encircled energy function (EEF). The Half Power Dimameter (HPD) and 90\% encircled energy (W90) can be extracted from the EEF, i.e.\ HPD~=~$\ang{;2.0;}$ and W90~=~$\ang{;5.6;}$. 

\begin{figure}[ht]%
        \begin{minipage}[b]{0.48\linewidth} 
            \centering 
            \includegraphics[width=1.05\textwidth,left]{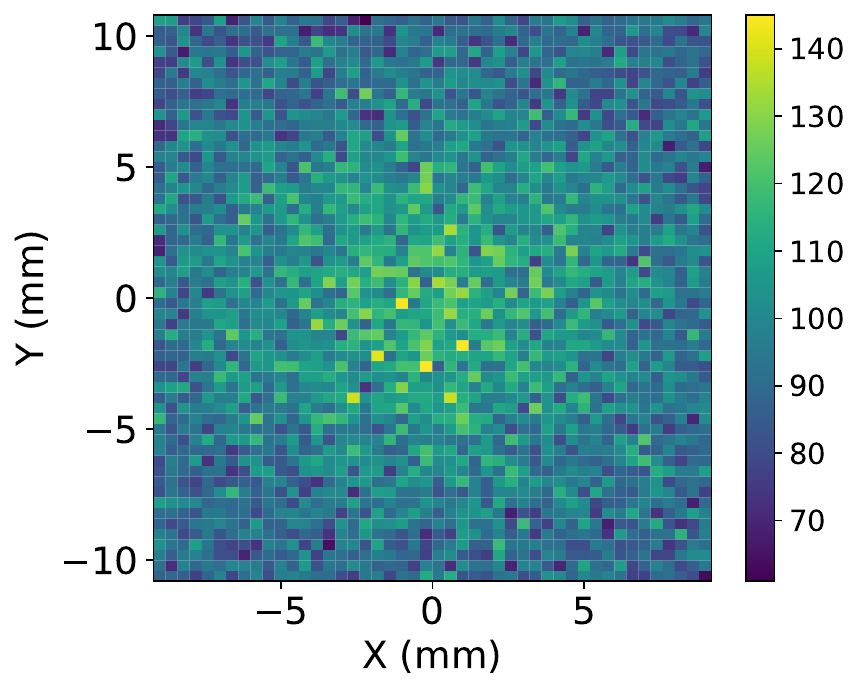} 
        \end{minipage} 
        \hspace{0.5cm}
        \begin{minipage}[b]{0.48\linewidth} 
            \centering 
            \includegraphics[width=1.02\textwidth]{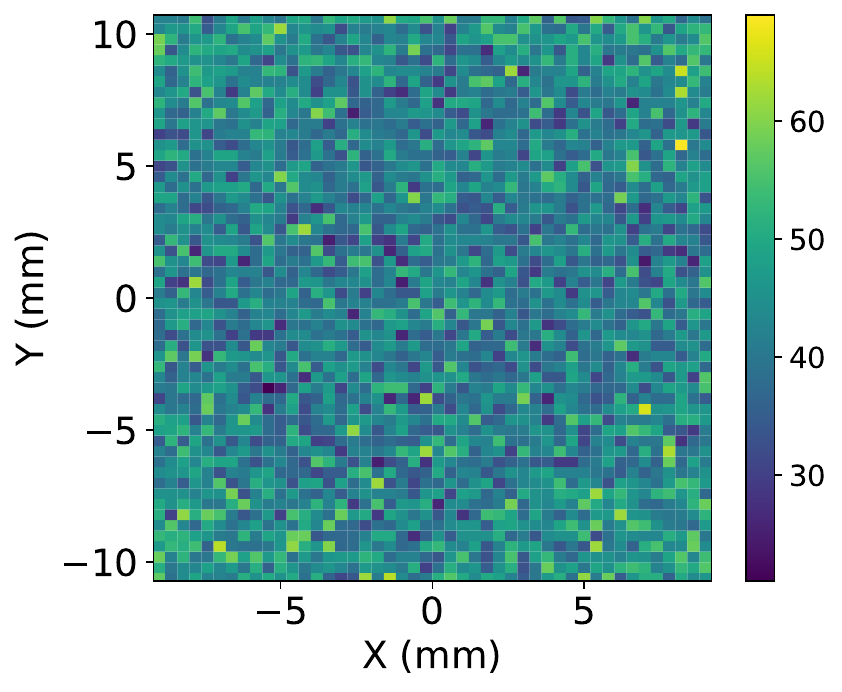} 
        \end{minipage}

        \vspace{0.3cm}
        \begin{minipage}[b]{1\linewidth} 
            \centering 
            \includegraphics[width=0.44\textwidth]{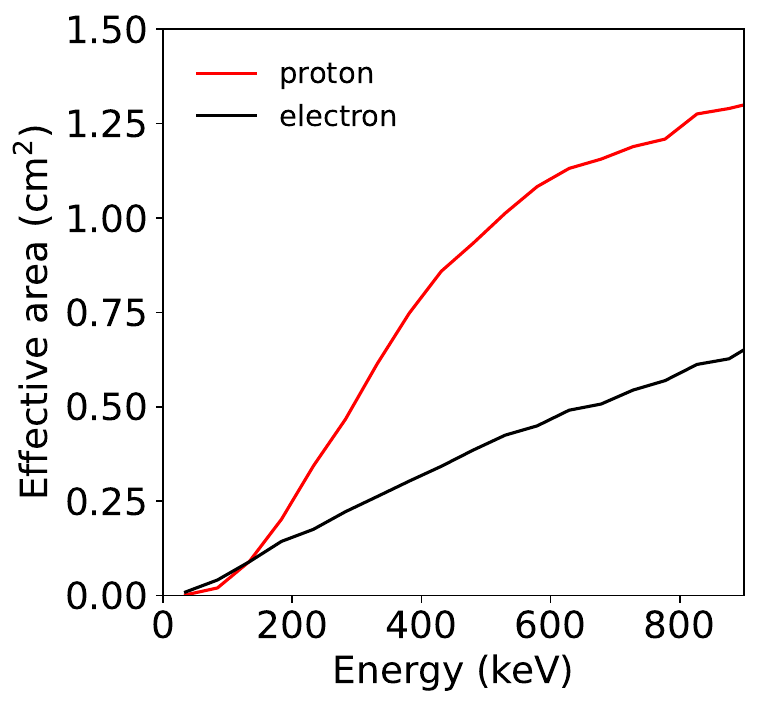} 
        \end{minipage}
        \vspace{0.05cm}
        \caption{Distribution of low-energy protons (upper-left panel) and electrons (upper-right panel) on the focal plane. Simulated on-axis effective area of low-energy protons and electrons through the focusing mirror of a type-B CATCH satellite, respectively (lower panel)}
        \label{PSFandEAcharged}
\end{figure}

The multiple scattering process (MSC) with a limited step-size is used to reproduce the funneling effect of low-energy charged particles through the Wolter-I type optics, i.e.\ G4WentzelVIModel for protons and G4GoudsmithSaundersonMscModel for electrons. It is shown that low-energy charged particles can leave the surface near the incident position of the Wolter-I mirrors~\cite{qi2020geant4}. The MSC model calculates multiple small-angle scatterings as one large-angle scattering. Thus, a step-size limitation is required in this case (i.e.\ grazing incidence) to reproduce the funneling effect. It can be seen in Figure~\ref{PSFandEAcharged} that the distribution of low-energy protons is slightly concentrated in the center while the distribution of low-energy electrons is relatively flat in the case of the focusing mirror of a type-B CATCH satellite. The corresponding on-axis effective area of the low-energy charged particles is plotted in the lower panel of Figure~\ref{PSFandEAcharged}.

\subsubsection{Detector}

\begin{figure}[ht]%
        \begin{minipage}[b]{0.48\linewidth} 
            \centering 
            \includegraphics[width=0.95\textwidth,left]{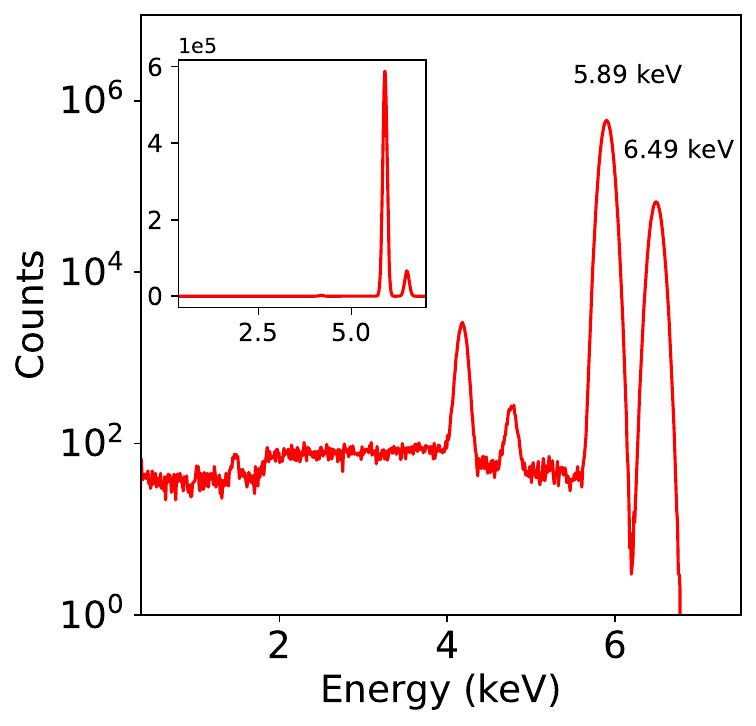} 
        \end{minipage}
        \hspace{0.3cm}
        \begin{minipage}[b]{0.48\linewidth} 
            \centering 
            \includegraphics[width=1\textwidth]{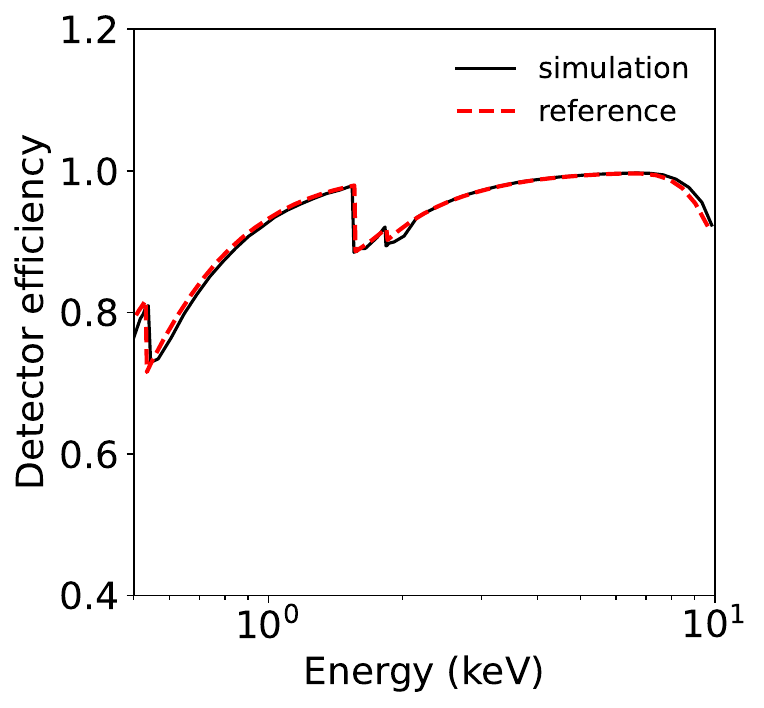} 
        \end{minipage} 
        \vspace{0.1cm}
        \caption{Simulated energy spectrum of the radioactive source $^{55}$Fe (left panel). Simulated total detection efficiency as a function of the energy in comparison with the reference (right panel)}
        \label{Fe55andefficiency}
\end{figure}

The depletion layer of the SDDs is defined as the sensitive region to the passage of particles. The class of SensitiveDetector is used to simulate the detector readout, including the energy deposited, position, and time information. The experimental energy resolution is used to re-sample the energy deposited as follows, 
\begin{equation}
    \textup{FWHM}=a\sqrt{b+\frac{cE}{a}},
\end{equation}
where $a$~=~38.71, $b$~=~1.54, $c$~=~73.06, and FWHM is in units of eV and $E$ is in units of keV. The left panel of Figure~\ref{Fe55andefficiency} shows the simulated energy spectrum of $^{55}$Fe. Typical structures are well reproduced, including the full-energy peak, escape peak, Compton platform and fluorescence peak. The right panel of Figure~\ref{Fe55andefficiency} shows the simulated total detection efficiency, which is consistent with the reference calculated using the analytical formula $e^{-\mu_{\textup{Al}}d_{\textup{Al}}}e^{-\mu_{\rm SiO_{2}}d_{\rm SiO_{2}}}(1-e^{-\mu_{\textup{Si}}d_{\textup{Si}}})$ with the attenuation coefficient from NIST~\footnote{http://physics.nist.gov/PhysRefData/FFast/html/form.html}. 

\subsection{Primaries generation}\label{3.3}

The primaries need to be initialized for the background simulation, including the definition of the type, energy, momentum direction, and position. The types of primaries and the corresponding energy spectra in the radiation environment are described in Section~\ref{sec2}. The energy can then be randomly sampled according to their spectra. In this section, the sampling of the momentum direction and position is presented. In a conventional background simulation, the satellite is assumed to be in an isotropic radiation environment. Considering the specific geometrical configuration of a type-B CATCH satellite, the background is simulated with the incident primaries both outside and inside the FoV of the telescope, respectively. It corresponds to two different strategies of the random sampling of the momentum direction and position. 

\subsubsection{Isotropic incidence outside the FoV}\label{sec3.3.1}

\begin{figure}[ht]%
        \begin{minipage}[b]{0.48\linewidth} 
            \centering 
            \includegraphics[width=0.75\textwidth]{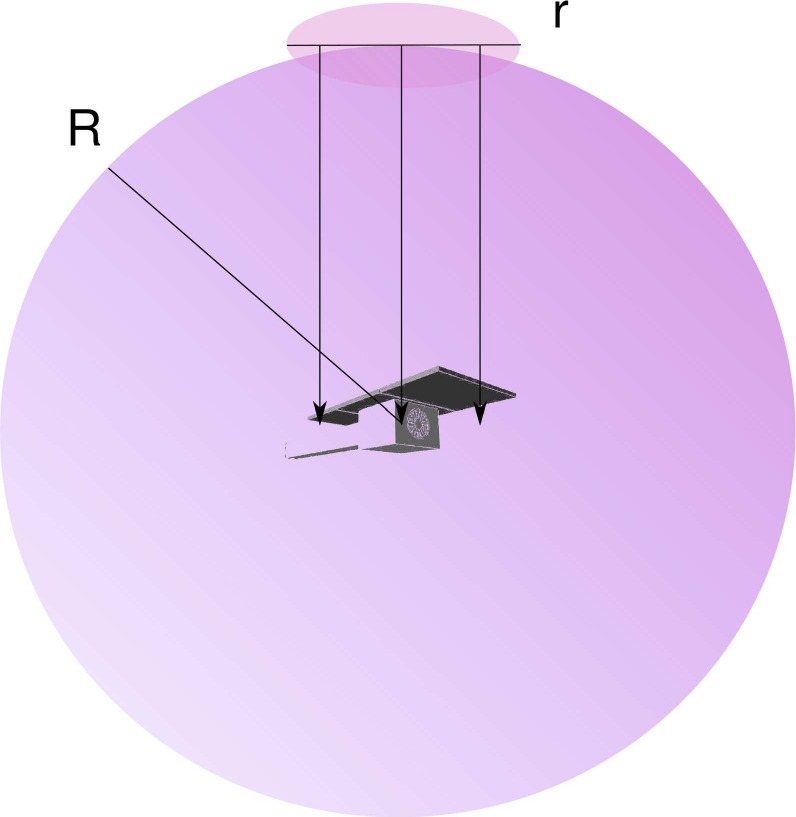} 
        \end{minipage} 
        \hspace{-0.2cm}
        \begin{minipage}[b]{0.48\linewidth} 
            \centering 
            \includegraphics[width=0.75\textwidth]{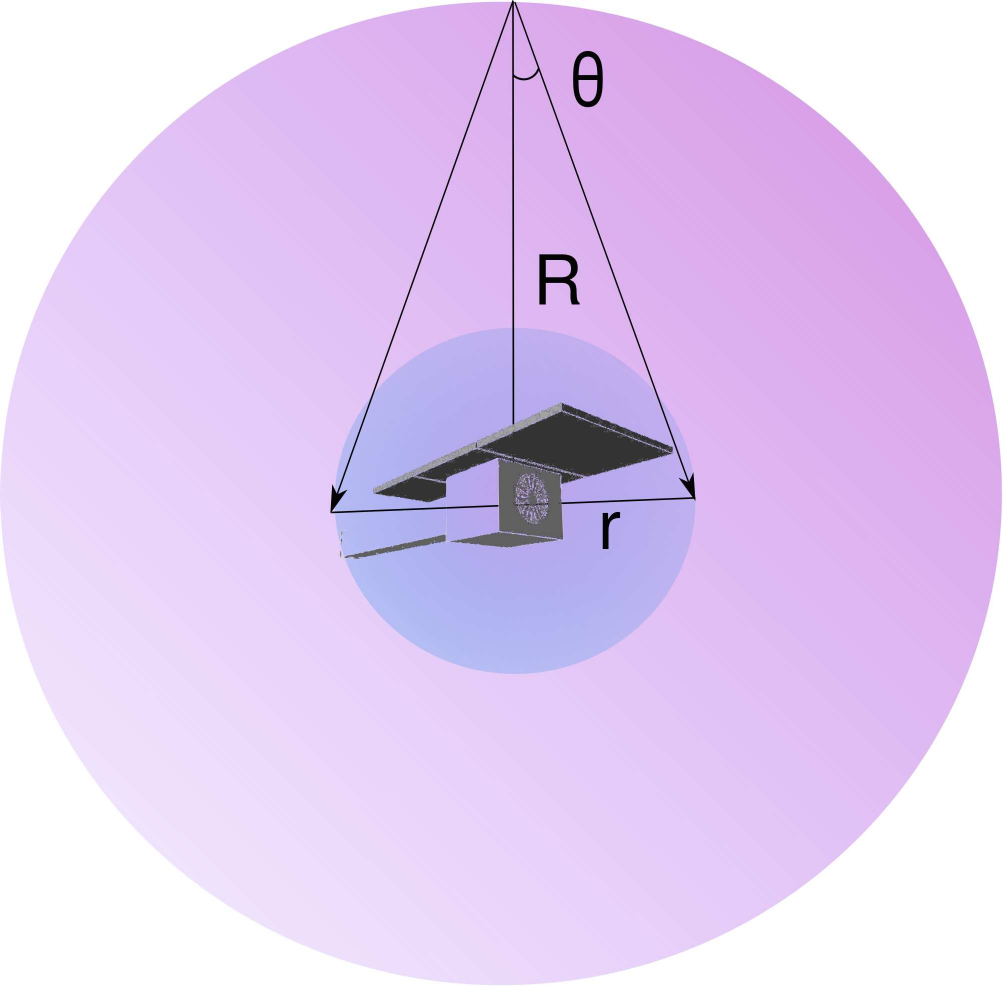} 
        \end{minipage}
        \vspace{0.6cm}
        \caption{Schematic view of the position and momentum direction sampling of primaries with isotropic incidence outside the FoV}
        \label{sample}
\end{figure}

A point is uniformly sampled on the surface of a sphere (with radius $R$). The negative position vector of this point is the momentum direction of primaries. The incident position of primaries is uniformly sampled within the tangent circle centering at this point (with radius $r$). The radius $r$ is chosen to cover the outline of the focal plane detector and its surrounding materials~\cite{zhang2022estimate} (see left panel of Figure~\ref{sample}). 

Another strategy is to uniformly sample a point on the surface of the sphere (with radius $R$). It represents the incident position of primaries. The momentum direction of primaries is biased with the cosine law. To increase the efficiency of the simulation, the emission angle $\theta$ is restricted by a maximum value $\theta_\textup{max}$\cite{campana2013background}. This value is determined by the size of the two spheres as follows~\cite{campana2013background}, 
\begin{equation}
    \theta_\textup{max}=\arctan\Big(\frac{r}{R}\Big),\label{equ:thetamax}
\end{equation}
where the smaller sphere (with radius $r$) covers the outline of the focal plane detector and its surrounding materials (see right panel of Figure~\ref{sample}). The two methods for sampling the momentum direction and position of incident primaries are tested. They produce the same background level with a proper choice of parameters.  

\subsubsection{Isotropic incidence inside the FoV}\label{sec3.3.2}

\begin{figure}[ht]%
        \centering 
        \includegraphics[width=0.6\textwidth]{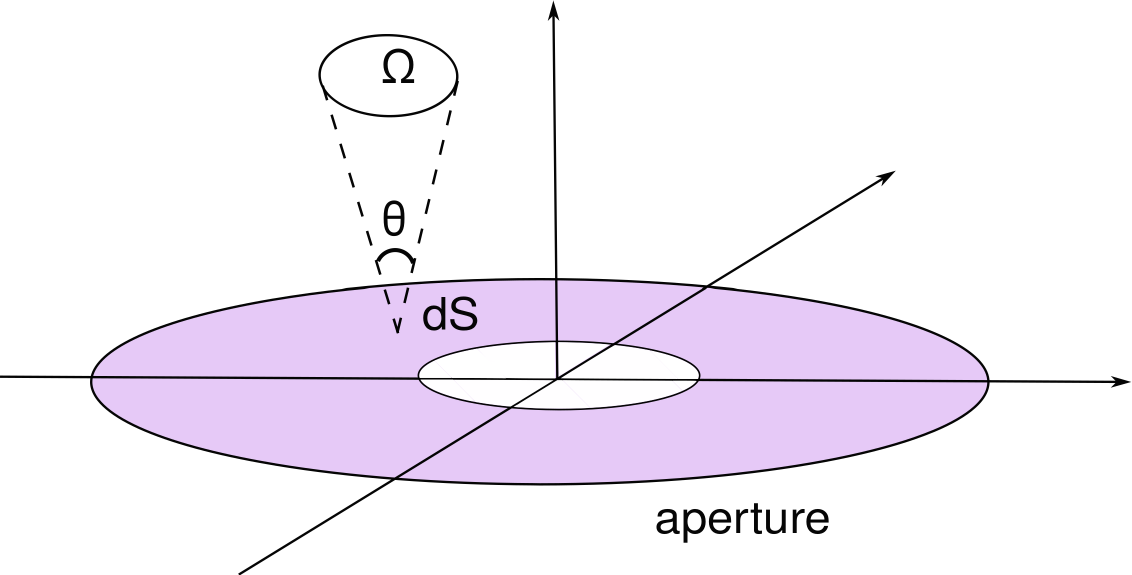}
        \vspace{0.55cm}
        \caption{Schematic view of the position and momentum direction sampling of primaries with isotropic incidence inside the FoV}
        \label{aperture}
\end{figure}

The X-rays and low-energy charged particles can reach the focal plane detectors through the Wolter-I focusing mirror~\cite{qi2020geant4}, which contributes to parts of the in-orbit background. The sampling methods discussed in Section~\ref{sec3.3.1} are not efficient for the simulation of this part of the background, because the FoV of the mirror is relatively small. Thus, primaries are uniformly sampled on the aperture of the telescope (circular ring shape) with the incident direction following a cosine law~\cite{zhao2013geometric} (see Figure~\ref{aperture}). 

It should be noted that the FoV discussed here is slightly different from its conventional use in astronomy. In this work, the FoV is defined as the maximum angular area, outside which the primaries rarely contribute to the in-orbit background through the Wolter-I focusing mirror. The FoV is different for each type of primaries, i.e.\ $\theta_\textup{max}$~=~$\ang{;90;}$ for X-rays, $\theta_\textup{max}$~=~$\ang{11;;}$ for protons, and $\theta_\textup{max}$~=~$\ang{15;;}$ for electrons~\cite{qi2021preliminary}.

\section{Results and discussions}

 The energy spectra described in Section~\ref{sec2} are used to sample the energy of primaries for the in-orbit background simulation of a type-B CATCH satellite. The momentum direction and incident position of primaries are sampled isotropically (Section~\ref{3.3}). The primaries interact with matters in a type-B CATCH satellite according to the mass modeling (Section~\ref{sec3.1}). The underlying interactions are described by the built-in physics libraries of Geant4 and the user-defined physics processes for the focusing X-ray telescopes (Section~\ref{3.2}). The energy deposited in a SDD as well as the information of the primary of the same event are recorded for further data analysis using ROOT. For example, the energy spectra of the SDDs induced by primaries outside and inside the FoV can be selected using the information of the primary incident position and momentum direction. As for the albedo $\gamma$-ray, the energy spectra of the SDDs are extracted using primaries within a certain solid angle due to the Earth's occlusion (assuming the satellite pointing to the zenith). The energy spectra of SDDs are then normalized with the time constant $t$,
\begin{equation}
t=\frac{N}{F\cdot S\cdot \Omega},
\end{equation}
where $F$ is the integral spectrum of incident particles, $S$ is the area of incident position, $\Omega$ is the solid angle of incident momentum direction. The normalized in-orbit background can be classified into two categories according to the FoV, i.e.\ the background induced by primaries inside and outside the FoV of the aperture, respectively.
 
\begin{figure}[ht]%
        \begin{minipage}[b]{1\linewidth} 
            \centering 
            \includegraphics[width=0.80\textwidth]{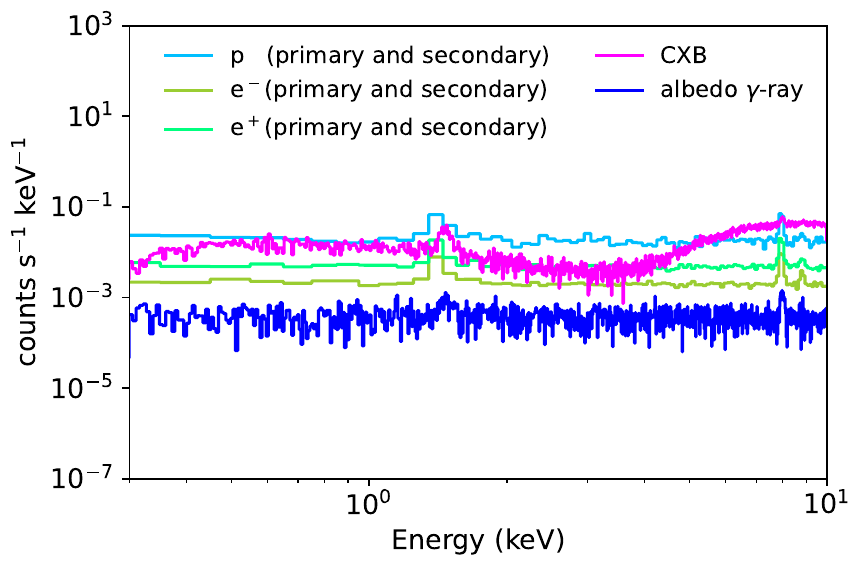} 
        \end{minipage} 
        \begin{minipage}[b]{1\linewidth} 
            \centering 
            \includegraphics[width=0.80\textwidth]{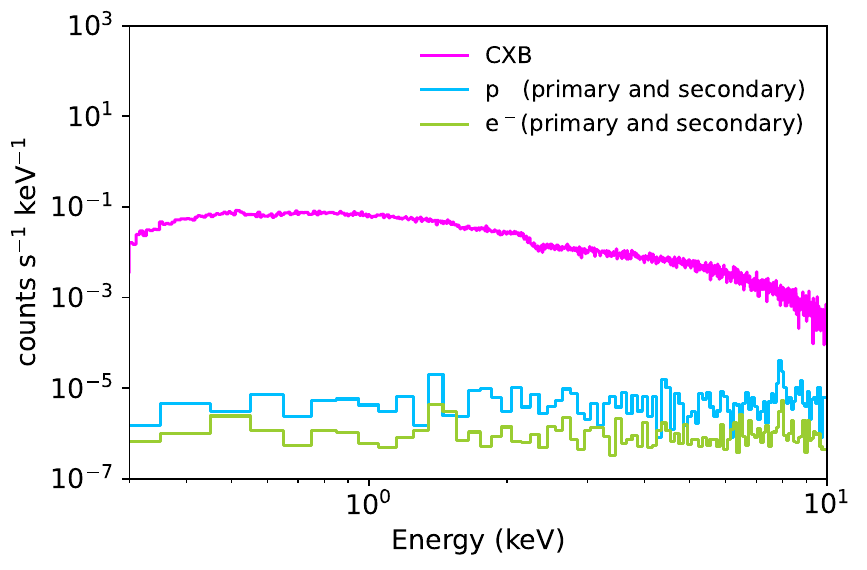} 
        \end{minipage}
        \vspace{0.1cm}
        \caption{Persistent in-orbit background induced by primaries outside the aperture (upper panel), and inside the aperture (lower panel)}
        \label{avergedbackground}
\end{figure}

The persistent in-orbit background is calculated considering the CXB, atmospheric albedo $\gamma$-ray, primary and secondary cosmic-ray particles in the space radiation environment. The normalized energy spectrum of the entire SDDs (4 modules, 256 pixels) induced by each component of the space radiation environment is plotted in  Figure~\ref{avergedbackground}. To clarify the origin, the energy spectra induced by primaries outside and inside the aperture are plotted in the upper and lower panels, respectively. In the upper panel, the energy spectra induced by protons, positrons, electrons, and albedo $\gamma$-ray have similar shapes, which contain a platform and a few fluorescence peaks. The fluorescence peaks stem from the shielding materials of Al and Cu (K-shell lines at 1.49~keV, 1.55~keV, 8.05~keV, 8.90~keV). The energy spectrum induced by CXB has a different shape with a bump above 4~keV; the possible causes are still under investigation. In the lower panel, the dominant component is CXB which can go through the focusing telescope and deposit energy in the SDDs (0.3--10~keV). Its shape is different from the one in the upper panel. It is mainly due to the instrumental response of the Wolter-I focusing mirror, e.g.\ the kink around 2~keV is consistent with the one in the effective area curve (see the right panel of Figure~\ref{RandEA}). The primary and secondary cosmic-ray particles can rarely penetrate through the Wolter-I focusing mirror and despite energy in the SDDs (0.3--10~keV), and the background level is four orders of magnitude lower than that in the upper panel. The integrated backgrounds in different energy bands are summarized in Table~\ref{averagedcountrates}. 

\begin{table}[ht]
\begin{center}
\begin{minipage}{1\textwidth}
\caption{Summary of the persistent in-orbit backgrounds of the entire SDD array, integrated in different energy ranges}\label{averagedcountrates}
\resizebox{\textwidth}{!}{
\begin{tabular*}{1.5\textwidth}{@{\extracolsep{\fill}}lcccccc@{\extracolsep{\fill}}}
\toprule%
& \multicolumn{3}{@{}c@{}}{Outside the aperture} & \multicolumn{3}{@{}c@{}}{Inside the aperture} \\\cmidrule{2-4}\cmidrule{5-7}%
Particle & 0.3-2~keV & 2-10~keV & 0.3-10~keV & 0.3-2~keV & 2-10~keV & 0.3-10~keV \\
\midrule
CXB  & $2.16\times 10^{-2}$ & $1.94\times 10^{-1}$ & $2.16\times 10^{-1}$ & $8.97\times 10^{-2}$ & $4.19\times 10^{-2}$ & $1.31\times 10^{-1}$\\
p  & $4.37\times 10^{-2}$ &  $1.51\times 10^{-1}$ & $1.93\times 10^{-1}$  & $9.80\times 10^{-6}$ & $4.72\times 10^{-5}$ & $5.67\times 10^{-5}$\\
$e^{-}$  & $4.68\times 10^{-3}$ &  $1.74\times 10^{-2}$ & $2.19\times 10^{-2}$  & $2.31\times 10^{-6}$ & $8.40\times 10^{-6}$ & $1.06\times 10^{-5}$\\
$e^{+}$  & $1.11\times 10^{-2}$ &  $4.14\times 10^{-2}$ & $5.20\times 10^{-2}$  & -- & -- & --\\
Albedo $\gamma$-ray  & $7.48\times 10^{-4}$ &  $3.12\times 10^{-3}$ & $3.86\times 10^{-3}$  & -- & -- & --\\
Total  & $8.18\times 10^{-2}$ &  $4.07\times 10^{-1}$ & $4.86\times 10^{-1}$  & $8.97\times 10^{-2}$ & $4.20\times 10^{-2}$ & $1.31\times 10^{-1}$\\
\botrule
\end{tabular*}}
\footnotetext{Note: The unit above is $\rm counts\:s^{-1}$, $S_\textup{detector}=3.74~\mathrm{cm^{2}}$}
\end{minipage}
\end{center}
\end{table}

\begin{figure}[ht]%
        \begin{minipage}[b]{1\linewidth} 
            \centering 
            \includegraphics[width=0.80\textwidth]{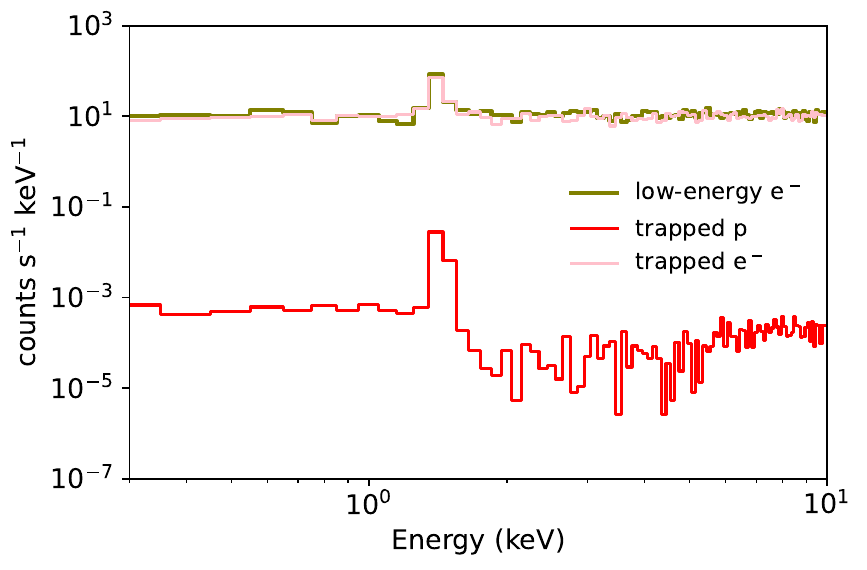} 
        \end{minipage} 
        \begin{minipage}[b]{1\linewidth} 
            \centering 
            \includegraphics[width=0.80\textwidth]{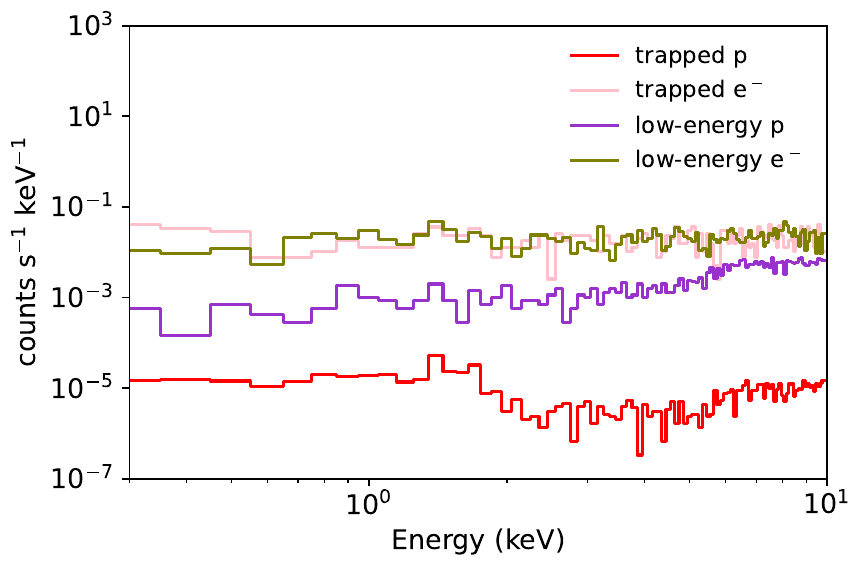} 
        \end{minipage}
        \vspace{0.1cm}
        \caption{Dynamic in-orbit background induced by primaries outside the aperture (upper panel), and inside the aperture (lower panel)}
        \label{variationbackground}
\end{figure}

The actual in-orbit background varies with time (i.e.\ position) and attitude rather than quiescence. For example, the flux of primary cosmic-ray particles changes periodically depending on the solar activity~\cite{gleeson1968solar}. The flux of charged particles increases in the South Atlantic Anomaly (SAA) region. The enhancement of low-energy charged particles is observed near the geomagnetic equator. For simplicity, the dynamic in-orbit background is calculated in this work considering the trapped charged particles by the Earth's radiation belts and low-energy charged particles near the geomagnetic equator. Figure~\ref{variationbackground} shows the normalized energy spectra of the entire SDDs induced by the dynamic components. To clarify the origin, the energy spectra induced by primaries outside and inside the aperture are plotted in the upper and lower panels, respectively. In the upper panel, the dominant components are electrons with the background level five orders of magnitude higher than that of protons. The fluorescence peaks from Al (K$_{\alpha}$ line at 1.49~keV and K$_{\beta}$ line at 1.55~keV) can be observed, while the fluorescence peaks from Cu (K$_{\alpha}$ line at 8.05~keV and K$_{\beta}$ line at 8.90~keV) is not observed. In the lower panel, the normalized energy spectra of electrons and protons are nearly flat. The funneling effect of low-energy charged particles (including charged particles in the radiation belts and near the geomagnetic equator) through the Wolter-I optics is much stronger than that of cosmic-ray particles. This is mainly because the considered energy range is much lower than those of cosmic-ray particles and the corresponding flux intensity is much higher. The integrated backgrounds in different energy bands are summarized in Table~\ref{timecountrates}. 

\begin{table}[ht]
\begin{center}
\begin{minipage}{1\textwidth}
\caption{Summary of the dynamic in-orbit backgrounds of the entire SDD array, integrated in different energy ranges}\label{timecountrates}
\resizebox{\textwidth}{!}{
\begin{tabular*}{1.5\textwidth}{@{\extracolsep{\fill}}lcccccc@{\extracolsep{\fill}}}
\toprule%
& \multicolumn{3}{@{}c@{}}{Outside the aperture} & \multicolumn{3}{@{}c@{}}{Inside the aperture} \\\cmidrule{2-4}\cmidrule{5-7}%
Particle type & 0.3-2~keV & 2-10~keV & 0.3-10~keV & 0.3-2~keV & 2-10~keV & 0.3-10~keV \\
\midrule
Low-energy p  & -- & -- & --  & $1.59\times 10^{-3}$ & $2.91\times 10^{-2}$ & $3.05\times 10^{-2}$\\
Low-energy $e^{-}$  & $2.78\times 10^{1}$ & $8.76\times 10^{1}$  & $1.14\times 10^{2}$  & $3.70\times 10^{-2}$ & $1.69\times 10^{-1}$ & $2.04\times 10^{-1}$\\
Trapped p  & $4.10\times 10^{-3}$ &  $1.06\times 10^{-3}$ & $5.16\times 10^{-3}$  & $3.25\times 10^{-5}$ & $5.11\times 10^{-5}$ & $8.33\times 10^{-5}$\\
Trapped $e^{-}$  & $2.53\times 10^{1}$ &  $7.96\times 10^{1}$ & $1.04\times 10^{2}$  & $3.70\times 10^{-2}$ & $1.58\times 10^{-1}$ & $1.94\times 10^{-1}$\\
Total  & $5.31\times 10^{1}$ &  $1.67\times 10^{2}$ & $2.18\times 10^{2}$  & $7.56\times 10^{-2}$ & $3.56\times 10^{-1}$ & $4.28\times 10^{-1}$\\
\botrule
\end{tabular*}}
\footnotetext{Note: The unit above is $\rm counts\:s^{-1}$, $S_\textup{detector}=3.74~\mathrm{cm^{2}}$}
\end{minipage}
\end{center}
\end{table}

\begin{figure}[ht]%
        \centering 
        \includegraphics[width=0.6\textwidth]{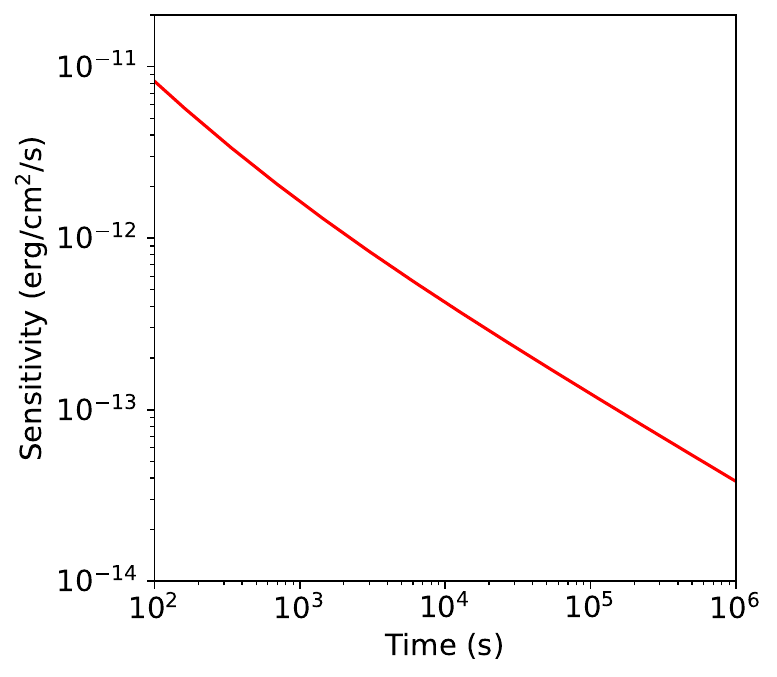}
        \vspace{0.55cm}
        \caption{Sensitivity of a type-B CATCH satellite as a function of the exposure time in the ideal case}
        \label{sensitivity}
\end{figure}

The simulated in-orbit background of a type-B CATCH satellite can be used to estimate the observation sensitivity of the instrument, optimize the instrument design, evaluate the onboard storage requirements, and infer the degradation of detector performances due to the radiation damage. As an example, the sensitivity of a type-B CATCH satellite is calculated using the simulated persistent in-orbit background. Considering the limited imaging capability of the optics of a type-B CATCH satellite, the focal spot of the target source (assuming point source) only covers a few pixels ($\approx$~7 pixels with respect to W90~=~$\ang{;5.6;}$). The in-orbit background within the focal spot is less than that of the entire SDDs (see Table~\ref{7pixels}). The observation sensitivity of a type-B CATCH satellite in the ideal case can be obtained from the following formula~\cite{vianello2018significance},
\begin{equation}
    \hat{M}=a+b \sqrt{B},
\end{equation}
where $a$~=~11.090, $b$~=~7.415 when considered 99\% detection efficiency above 5$\sigma$, $\hat{M}$ and $B$ is in units of counts. With an exposure time of $10^{4}$ and $10^{5}$ seconds for a Crab-like source spectrum (the model is phabs*powerlaw in XSPEC~\cite{zhang2022estimate}), a type-B CATCH satellite is able to achieve a sensitivity of $4.22\times 10^{-13}$~$\rm erg\:cm^{-2}\:s^{-1}$ and $1.24\times 10^{-13}$~$\rm erg\:cm^{-2}\:s^{-1}$ in the energy band 0.3--10~keV, respectively. The observation sensitivity as a function of the exposure time is plotted in Figure~\ref{sensitivity}. 

To improve the observation sensitivity, the persistent in-orbit background can be decreased with an optimized shielding design. A type-B CATCH satellite has specific geometrical configurations, i.e.\ the focal plane detector is deployed 2 meters away from the platform cabin without massive materials surrounding it. As a consequence, the major contribution of the persistent in-orbit background comes from the CXB and the cosmic-ray protons (primary and secondary). To decrease corresponding background levels, the configurations of the Al and Cu layers around the focal plane detector can be re-designed. For example, the height and thickness of the Al layer can be increased appropriately to shield particles outside the aperture, an Al slab can be added to the back of the detector to suppress the fluorescence from Cu. A detailed study to improve the shield design and decrease the background will be performed in the future. Additionally, a magnetic diverter is usually equipped in Wolter-I type focusing telescopes to decrease the background of low-energy charged particles through the aperture, e.g.\ SIMBOL-X~\cite{spiga2008magnetic}, ATHENA~\cite{riva2018study}, and EP~\cite{wang2020design}. In terms of the persistent background of a type-B CATCH satellite, the charged particles through the aperture have a negligible contribution. In terms of the dynamic background of a type-B CATCH satellite, the contribution of the charged particles through the aperture is three orders of magnitude lower than that outside the aperture. This suggests that the magnetic diverter just underneath the optics may be unnecessary in this kind of micro-satellites.

\begin{table}[ht]
\begin{center}
\begin{minipage}{1\textwidth}
\caption{Persistent in-orbit background within the focal spot (7 pixels in the SDD array), integrated in the energy band 0.3--10~keV}\label{7pixels}%
\resizebox{\textwidth}{!}{
\begin{tabular}{@{}lccc@{}}
\toprule
Particle type & Outside the aperture & Inside the aperture & Total \\
\midrule
CXB   &  $3.86\times 10^{-3}$   & $5.05\times 10^{-3}$ &  $8.91\times 10^{-3}$ \\
p   &  $5.50\times 10^{-3}$   & $1.01\times 10^{-6}$ &  $5.50\times 10^{-3}$ \\
$e^{-}$  & $6.06\times 10^{-4}$ &  $2.29\times 10^{-7}$ & $6.06\times 10^{-4}$ \\
$e^{+}$  & $1.42\times 10^{-3}$ &  -- & $1.42\times 10^{-3}$ \\
Albedo $\gamma$-ray  & $1.12\times 10^{-4}$ &  -- & $1.12\times 10^{-4}$ \\
Total  & $1.15\times 10^{-2}$ &  $5.05\times 10^{-3}$ & $1.65\times 10^{-2}$ \\
\botrule
\end{tabular}}
\footnotetext{Note: The unit above is $\rm counts\:s^{-1}$, and the number of the pixels is chosen according to the W90 of the optics}
\end{minipage}
\end{center}
\end{table}

\section{Summaries and conclusions}

CATCH is an intelligent X-ray satellite constellation mission designed to study the dynamic universe. The constellation includes three types of micro-satellites (A, B, and C). The type-B CATCH satellites are mainly used for locating the transients, and performing the spectral and timing measurements in the energy band between 0.3~keV and 10~keV with high sensitivities. To achieve its scientific goals, a type-B satellite is equipped with a lightweight Wolter-I focusing mirror and an array of position-sensitive multi-pixel SDDs. Such a micro-satellite has a specific geometrical configuration, i.e.\ the focal plane detector is deployed 2 meters away from the platform cabin without massive materials surrounding it.

In this work, the in-orbit background simulation is performed for a type-B CATCH satellite using the Geant4 toolkit. The geometry model is built including the platform, Wolter-I optics, SDD array and detector shielding. The physics model is established and verified to adapt to the application of X-ray telescopes, namely the total internal reflection of X-rays on the mirror surface and the funneling effect of low-energy charged particles through the optics. The energy, momentum direction, and position of each type of primaries are sampled according to their own characteristics in the space radiation environment. The simulation shows that the persistent in-orbit background is dominated by the CXB and the cosmic-ray protons, $5.40\times 10^{-1}$~$\rm counts\:s^{-1}$ out of $6.17\times 10^{-1}$~$\rm counts\:s^{-1}$. The dynamic in-orbit background is estimated taking into account the trapped charged particles in the radiation belts and low-energy charged particles near the geomagnetic equator. It is dominated by the electrons outside the aperture, $2.180\times 10^{2}$~$\rm counts\:s^{-1}$ out of $2.184\times 10^{2}$~$\rm counts\:s^{-1}$. 

The simulated in-orbit background of a type-B CATCH satellite can be used to estimate the observation sensitivity of the instrument, optimize the instrument design, evaluate the onboard storage requirements, and infer the degradation of detector performances due to the radiation damage. As an example, the observation sensitivity within the focal spot is estimated using the simulated persistent in-orbit background, which is $4.22\times 10^{-13}$~$\rm erg\:cm^{-2}\:s^{-1}$ with an exposure time of 10$^{4}$~s and a Crab-like source spectrum. To achieve a better observation sensitivity, the shielding design can be optimized, e.g.\ increasing the height and thickness of the Al layer and adding an Al slab at the bottom of the detector. Additionally, it is suggested that the magnetic diverter just underneath the optics may be unnecessary in this kind of micro-satellites, because the dynamic background induced by low-charged particles outside the aperture is around 3 orders of magnitude larger than that inside the aperture. 
 
\section*{Acknowledgments}

We would like to appreciate the CATCH collaboration. We acknowledge the support from the National Natural Science Foundation of China (NSFC), Grant Nos. 12122306 and 12003037, the Strategic Priority Research Program of the Chinese Academy of Sciences XDA15016400, the CAS Pioneer Hundred Talent Program Y8291130K2. We also acknowledge the Scientific and technological innovation project of IHEP E15456U2.

\section*{Author Contributions}

Jingyu Xiao and Liqiang Qi wrote the main manuscript. Shuang-Nan Zhang, Lian Tao, and Zhengwei Li assisted with the critical revision of the article. Juan Zhang verified the analytical methods. All authors reviewed the manuscript and contributed to the development of the in-orbit background simulation of CATCH.

\section*{Funding}

We acknowledge funding support from the National Natural Science Foundation of China
(NSFC) under grant Nos. 12122306 and 12003037, the Strategic Priority Research Program of the Chinese
Academy of Sciences XDA15016400, the CAS Pioneer Hundred Talent Program Y8291130K2, and the Scientific and technological innovation project of IHEP
E15456U2.

\section*{Conflict of interest}

The authors declare that they have no conflict of interest.


\bibliography{sn-bibliography}


\end{document}